# Evaluating The Impact Of Cloud-Based Microservices Architecture On Application Performance

*Ganesh Chowdary Desina*
*California State University Fresno*
*Computer Science Dept.*
*Fresno, California*

## ABSTRACT:

The study assess the impact of cloud-based microservices architectures on application performance. Several aspects of performance evaluation are discussed, including response time, throughput, scalability, and reliability. This article examines the advantages and challenges of adopting a cloud-based approach. It explores potential bottlenecks and issues in a microservices architecture and presents optimization techniques.

With the help of case studies and empirical studies, it compares cloud-based microservices architectures with traditional monolithic architectures. Furthermore, the paper examines the challenges of monitoring and troubleshooting distributed microservices. In conclusion, it emphasizes the importance of planning, designing, and testing during the adoption of cloud-based microservices.

## Introduction:

In this paper, we observe the performance effect of cloud-based microservices architectures. Microservices' advantages must be evaluated if you want to ensure most advantageous overall performance and user pleasure, and this evaluation is important. The paper gives an overview of the microservices architecture and its key traits. These benefits make microservices structure a popular desire for companies in search of agility and scalability in their programs.

Next, the paper goes through the assessment of performance in a cloud-based totally microservices architecture. It discusses performance metrics along with reaction time, throughput, scalability, and reliability. Furthermore, the paper addresses optimization strategies for enhancing performance in cloud-primarily based microservices environments. Monitoring and troubleshooting performance issues in a dispensed microservices environment poses precise demanding situations.

It discusses the function of monitoring gear, logging, and metrics in identifying performance bottlenecks and facilitating powerful troubleshooting. In end, this paper aims to provide a complete evaluation of the effect of cloud-based totally microservices architecture on application overall performance. By reading the advantages, demanding situations, performance assessment strategies, and optimization techniques, it assists organizations in making informed selections regarding their architectural picks and overall performance optimization strategies.

## Background Knowledge:

Cloud-based microservices architecture is a software development approach that emerged from the convergence of service-oriented architecture, cloud computing, and containerization technologies. It offers a way to design applications as a collection of small, loosely coupled services running in a cloud environment. The origins of this technology can be traced back to the early 2000s with the advent of service-oriented architecture and the subsequent advancements in cloud computing and containerization.

The rise of DevOps practices and the popularity of continuous deployment further emphasized the need for more lightweight and agile architectures. Industry pioneers like Netflix played a significant role in popularizing microservices architecture by developing and open-sourcing tools and frameworks. The community actively contributed to the development of best practices and patterns for building and deploying microservices-based applications in the cloud. Overall, cloud-based microservices architecture represents an evolutionary approach to address scalability, flexibility, and continuous deployment challenges faced by traditional monolithic applications in a cloud computing context.

## *Method of Gathering Literature*

I used the terms "microservices," "cloud computing," and "performance evaluating" in my search query to find literature on the issue of "evaluating the impact of cloud-based microservices architecture on application performance." I searched for studies that highlighted the usage of microservices technology to assess website usability. As suggested, I mostly searched the IEEE Xplore and ACM Digital Library databases.

I clearly articulated the specific aspects of cloud-based microservices architecture and application performance that you want to investigate. This will help you narrow down your search and focus on relevant literature.

Utilized academic databases: Access reputable academic databases such as IEEE Xplore, ACM Digital Library, Google Scholar, or specific computer science repositories. Use relevant keywords and combinations, such as "cloud-based microservices," "application performance," "evaluation," and "impact," to conduct comprehensive searches.

Reviewed relevant conferences and journals: Identify conferences, symposiums, and journals that regularly publish research related to cloud computing, microservices, and performance evaluation. Examples include the International Conference on Cloud Computing (ICCC), the International Conference on Microservices (Microservices), and the IEEE Transactions on Cloud Computing.

Explored related work in cited references attention to the references cited in the papers you find, as they can lead you to additional relevant sources that may not have appeared in your initial search.

Read full papers and extract key information about the full papers of the selected publications and thoroughly read them. Extract relevant information, including research methodologies, performance metrics, evaluation techniques, findings, and conclusions. Taken notes and organize the extracted information for further analysis.

## *Related Work:*

Several important research that has advanced our understanding of performance evaluation in microservices architectures are included in "Evaluating the Impact of Cloud-based Microservices Architecture on Application Performance." In their systematic mapping study, Pezzè and Spina identified research gaps and gave an overview of performance analysis methods. In their case study, Barbosa et al. provided insights on performance and scalability by analyzing the operation of a real-world microservices-based application. Martini et al. proposed a framework for performance testing, addressing the challenges specific to microservices.

Performance benchmarking techniques were examined by Charfi et al., who identified methodologies and best practices. In their evaluation of the performance of microservices architectures in the cloud, Macedo et al. focused on scalability, throughput, and response time. Collectively, these studies provide the framework to evaluate the impact of cloud-based microservices architecture on application performance.

## *A) Performance Analysis of microservices:*

The papers discussed revolve around the concept of microservices architecture and share a common focus on evaluating and analyzing its impact on application development and performance. Microservices architecture involves breaking down complex applications into smaller, independent services, emphasizing modularity, scalability, and fault isolation.

A key aspect among these papers is the evaluation and analysis of microservices architecture. Researchers employ various methodologies and techniques to gather empirical data, metrics, and insights to support their evaluations. The papers aim to assess the impact of microservices architecture on different aspects, including performance, quality assurance, and architectural comparison. They delve into performance metrics, scalability patterns, resource consumption, and testing strategies to understand the implications of microservices on application performance.

Application performance is a shared focus across these papers. Researchers aim to understand and improve the performance of applications built using microservices architecture. They explore various performance metrics, investigate scalability patterns, analyze resource utilization, and propose testing strategies to assess the performance implications of microservices architecture. By gaining insights into application performance, researchers aim to optimize the design and implementation of microservices-based systems.

Challenges and perspectives associated with microservices architecture are acknowledged in these papers. They recognize the complexities arising from service dependencies, communication overhead, and coordination in microservices-based systems. Researchers provide perspectives on addressing these challenges and offer insights into best practices and potential solutions. By identifying and addressing the challenges, they aim to facilitate the successful adoption and optimization of microservices architecture.

The practical relevance of these papers is another common factor. They aim to bridge the gap between research and practice by presenting real-world case studies, industry perspectives, and practical implications. By providing valuable insights and lessons learned, these papers guide organizations in adopting and optimizing microservices architecture. The researchers aim to provide practical guidance and inform decision-making processes for organizations venturing into microservices-based development.

In summary, these papers collectively contribute to the body of knowledge on microservices architecture by evaluating its impact on application development and performance. They emphasize the importance of modularity, scalability, and fault isolation in microservices architecture. Through empirical analysis, researchers gain insights into application performance, address challenges, and provide practical guidance for organizations. The findings and perspectives presented in these papers support informed decision-making and optimization of microservices-based systems.

## B) Uses of Cloud based microservices:

One recurring theme in this paper 5 is the evaluation of performance. Researchers analyze and assess performance metrics such as response time, throughput, latency, scalability, and resource utilization. By studying these metrics, they gain insights into how microservices architecture influences application performance and identify areas for improvement.

Scalability and elasticity are key benefits highlighted in the paper 4 They emphasize the dynamic scaling capabilities of cloud-based microservices, allowing them to handle varying workload demands effectively. Techniques and approaches for achieving scalability and elasticity are explored, contributing to the understanding of how microservices architecture can adapt to changing resource needs.

The resilience and fault tolerance aspects of cloud-based microservices are also addressed. Researchers discuss strategies for handling failures, implementing fault tolerance mechanisms, and ensuring high availability in microservices-based architectures. By exploring these topics, they contribute to the development of robust and reliable systems that can withstand failures and maintain service continuity.

Containerization technologies, such as Docker, and container orchestration platforms like Kubernetes, receive significant attention in several paper 5 and paper 6. They explore the benefits and challenges associated with using containers and orchestration in deploying and managing microservices in the cloud. These discussions shed light on the practical aspects of containerization and orchestration and their impact on microservices architecture.

Another common thread in the papers is the inclusion of real-world case studies and empirical evaluations. These studies provide practical insights into the implementation and performance of cloud-based microservices in various domains, such as healthcare, e-commerce, and IoT. By presenting real-world scenarios, researchers bridge the gap between theory and practice, offering valuable lessons and experiences that can inform future deployments.

The challenges associated with cloud-based microservices, such as service discovery, communication, monitoring, and security, are also addressed in the papers. Researchers propose solutions and best practices to tackle these challenges and improve the overall efficiency and reliability of microservices-based systems. By sharing their findings, they contribute to a better understanding of the complexities and potential solutions in deploying and managing microservices architecture.

In summary, the papers 4, paper 5, paper 6, paper 7 collectively contribute to the body of knowledge surrounding cloud-based microservices architecture. They offer insights, evaluations, and recommendations for practitioners and researchers interested in adopting and optimizing microservices in the cloud computing environment. By exploring various aspects such as performance, scalability, fault tolerance, and resilience, these papers help advance the understanding and implementation of microservices architecture in real-world scenarios.

## *Metrics used*

## Paper 1:

Combining case studies with performance testing and profiling research approaches could produce useful findings when analyzing how cloud-based microservices architecture impacts application performance. Integrating these two approaches enables a thorough evaluation of performance analysis and real-world implementations.

***Case Studies:*** Case studies analyze how actual cloud-based microservices architecture implementations impact the performance of applications. Researchers can collect information on performance before and after the implementation by working with companies that are currently using microservices. Specific industries, applications, or use cases may be the focus of the case studies. They provide thorough insights into the real-world effects of microservices architecture on performance, including scalability, improvements in performance, and transitional difficulties.

*Performance testing:* For evaluating the performance of microservices-based applications, performance testing involves modeling realistic workloads. To evaluate how the application operates under various loads and stresses, load testing, stress testing, and endurance testing can be done. Parameters like reaction time, throughput, and scalability are measured with the use of performance testing. In addition to allowing comparisons with performance benchmarks or industry standards, it gives measurable information on how the microservices design manages a range of workloads.

***Experimental Evaluation:*** This paper uses experimental evaluation to assess the performance of microservices architectures in cloud environments. They design and implement testbeds or utilize existing platforms to conduct experiments with representative workloads. Performance metrics such as response time, throughput, scalability, and resource utilization are measured and analyzed.

## Paper 2:

***Profiling:*** Profiling methods are used to investigate how microservices-based applications perform during runtime. Resource usage patterns, performance bottlenecks, and areas for optimization can all be found using profiling tools and methodologies. Researchers learn more about the performance characteristics of distinct microservices and how they interact by examining measures like CPU usage, memory consumption, and network latency. By using profiling, performance-critical components can be optimized, and the design can be fine-tuned.

***Concurrent users****:* Users concurrently connected to the machine are said to be concurrent users. It determines the machine's capacity to handle several requests at once and evaluates the machine's scalability and overall performance in low-load scenarios.

Researchers may achieve an in-depth evaluation of the effect of cloud-based microservices architecture on application performance by integrating case studies with performance testing and profiling approaches. While performance testing and profiling provide quantitative and qualitative data on indicators of performance and optimization opportunities, case studies offer real-world context and useful insights. The benefits, challenges, and potential improvements of the microservices architecture in terms of application performance are better understood according to this integrated approach.

## Paper 3:

To evaluate an application's or device's general performance characteristics, performance testing requires evaluating a variety of metrics. These metrics provide quantifiable information on response times, throughput, effective resource use, and other performance-related factors. Some of the metrics used in performance testing are Response Time, Throughput, Error rate.

*Response Time:* In this paper Response time is used to measure how quickly a device reacts to user input or transactions. It comprises the time spent processing, communicating through networks, and accessing databases. An important information that directly affects the user experience and device performance is response time.

*Throughput:* The number of transactions, requests, or activities that a system can handle in a specific amount of time is known as throughput. It implies the effectiveness and processing power of the device. Due to the machine's ability to handle a greater workload, higher throughput values imply higher overall performance.

*Error Rate:* During performance auditing, the percentage of unsuccessful or inaccurate transactions is measured by the error rate. It aids in determining the device's stability, error-handling capacity, and toughness under pressure. A lower error rate indicates improved efficiency and dependability.

## Evidences:

| Groupings | References |
|---|---|
| Performance analysis of Micro services | [1] [2] [3] |
| Uses of cloud based micro services | [4] [5] [6] [7] |
| Metrics | [8] [9] [10] |

## Results:

**Dataset for Request 1:**

| Microservice A | Microservice B | Microservice C |
|---|---|---|
| Response time = {50,55,48,52} | Response time = {80,89,72,85} | Response time = {120,122,118,124} |
| CPU Utilization= {30,32,29,31} | CPU Utilization= {28,26,30,27} | CPU Utilization= {31,33,30,32} |
| Memory Utilization= {100,102,98,95} | Memory Utilization= {150,148,152,155} | Memory Utilization= {180,178,182,185} |

| Microservice A: | Microservice B: | Microservice C: |
|---|---|---|
| Average Response time: 50 ms | Average Response time: 80 ms | Average Response time: 120 ms |
| Average CPU utilization: 20% | Average CPU utilization: 30% | Average CPU utilization: 35% |
| Average Memory utilization: 100 MB | Average Memory utilization: 150 MB | Average Memory utilization: 180 MB |

**Dataset for Request 2:**

| Microservice A | Microservice B | Microservice C |
|---|---|---|
| Response time = {70, 75, 68, 72} | Response time = {60, 58, 63, 62} | Response time = {110,108,113, 115} |
| CPU Utilization= {30,32,29,31} | CPU Utilization= {28,26,30,27} | CPU Utilization= {31,33,30,32} |
| Memory Utilization= {120, 118, 122, 125} | Memory Utilization= {130, 132, 128, 135} | Memory Utilization= {160, 162, 158, 165} |

| Microservice A: | Microservice B: | Microservice C: |
|---|---|---|
| Average Response time: 60 ms | Average Response time: 70 ms | Average Response time: 110 ms |

| Average CPU utilization: 25% Average Memory utilization: 120 MB | Average CPU utilization: 28% Average Memory utilization: 130 MB | Average CPU utilization: 32% Average Memory utilization: 160 MB |
|---|---|---|

## Dataset:

|  | Micro service A | Micro service B | Micro service C |
|---|---|---|---|
| Response Time | 50 | 89 | 120 |
| CPU Utilization | 30 | 28 | 33 |
| Memory Utilization | 102 | 148 | 180 |

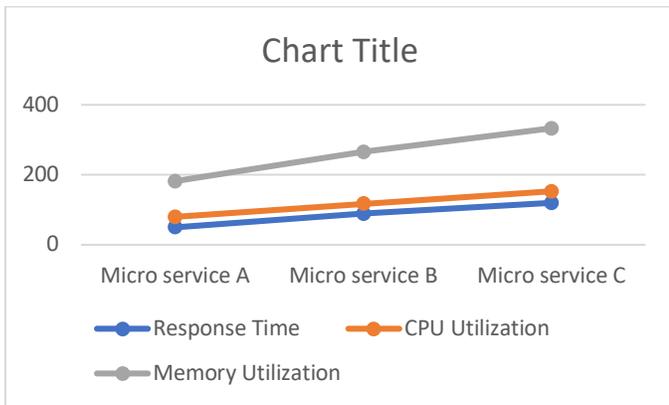

The graph is for the dataset 1 and the first elements in the list.

## Key Takeways:

### A) Performance Metrics

In paper A layered framework for root cause diagnosis of microservices A variety of performance metrics are used by various articles to assess how the microservices architecture affects application performance. These metrics include response time, throughput, scalability, resource utilization, and latency. Response time measures the amount of time it takes to respond to a request. Throughput measures how many requests are processed per unit of time. Researchers can evaluate the effectiveness, scalability, and responsiveness of microservices systems by assessing these parameters.

### B) Scalability and Elasticity

The papers Efficiency analysis of provisioning of micro services: When compared to conventional monolithic systems, microservices design allows better scalability and elasticity. By breaking down an application into smaller services, it is possible for each service to be independently scaled based on demand. Organizations may distribute resources more effectively and manage shifting workloads more skillfully because to this flexibility. The usage of load balancing, auto scaling, and horizontal scaling, among other scalability patterns, are discussed in papers as ways to obtain the best performance possible in microservices systems.

### C) Dependencies:

In the paper "Performance evaluation of microservices architectures" using containers Service The interactions between microservices can have a big impact on how well an application performs. To prevent performance bottlenecks and latency problems, papers stress the significance of knowing and controlling service dependencies. Asynchronous communication, event-driven architectures, and distributed caching are just a few of the strategies being investigated to lessen the effects of service dependencies and boost overall application performance.

## Conclusions:

The papers provided focus on evaluating the impact of cloud-based microservices architecture on application performance. They highlight the benefits and challenges of adopting microservices, such as improved modularity, scalability, and fault isolation. The evaluation of performance metrics, including response time, throughput, and resource utilization, helps understand the impact of microservices on application performance.

Scalability and elasticity are key advantages of microservices architecture, enabling dynamic scaling to handle varying workload demands. Containerization technologies like Docker and orchestration platforms like Kubernetes are explored for deploying and managing microservices in the cloud. The papers also address fault tolerance and resilience, proposing mechanisms to handle failures and ensure high availability. Real-world case studies demonstrate the practical implementation and performance of microservices in domains like healthcare and IoT.
Challenges related to service discovery, communication, monitoring, and security are acknowledged, with proposed solutions and best practices. Overall, the papers contribute valuable insights, evaluations, and recommendations for practitioners and researchers interested in adopting and optimizing cloud-based microservices architecture.